\documentstyle[12pt]{article}
\def\tfrac#1#2{{\textstyle{{#1}\over{#2}}}}
\def\half{\tfrac{1}{2}}
\begin{document}

\begin{titlepage}

\begin{flushright}
RAL-TR-97-069\\
%hep-ph/97???\\
\end{flushright}

\begin{center}
\baselineskip 24pt
{\Large {\bf Standard Model with Duality: Theoretical Basis}}\footnote{Review 
   talk given by the second author at the Cracow Summer School on Theoretical
   Physics held in May-June 1997 at Zakopane, Poland, to appear in Acta
   Physica Polonica.}\\
\vspace{.5cm}
\baselineskip 16pt 
{\large CHAN Hong-Mo}\\
chanhm\,@\,v2.rl.ac.uk\\
{\it Rutherford Appleton Laboratory,\\
  Chilton, Didcot, Oxon, OX11 0QX, United Kingdom}\\
\vspace{.2cm}
{\large TSOU Sheung Tsun}\\
tsou\,@\,maths.ox.ac.uk\\
{\it Mathematical Institute, University of Oxford,\\
  24-29 St. Giles', Oxford, OX1 3LB, United Kingdom}\\
\end{center}

\vspace{.3cm}
\begin{abstract}
The Dualized Standard Model which has a number of very interesting physical 
consequences is itself based on the concept of a nonabelian generalization
to electric-magnetic duality.  This paper explains first the reasons why 
the ordinary (Hodge) * does not give duality for the nonabelian theory 
and then reviews the steps by which these difficulties are surmounted, 
leading to a generalized duality transform formulated in loop space.  The 
significance of this in relation to the Dualized Standard Model is explained, 
and possibly also to some other areas.
\end{abstract}

\end{titlepage}

\clearpage

\section{Introduction}

From the standpoint of our present understanding and observation, the
Standard Model seems to encapsulate the major points of our knowledge in
particle physics but yet leaves many of its own ingredients unexplained.
Of these, the most striking are the origins of Higgs fields and fermion
generations.  Nor are details such as the fermion mass hierarchy or the CKM 
(Cabibbo--Kobayashi--Maskawa) mixing matrix \cite{ckm} given any theoretical 
explanations.  A way to further our understanding is perhaps to study more 
closely Yang--Mills theory itself, on which the Standard Model is based.  
Indeed, it was shown that by combining a recently derived generalized 
electric-magnetic duality for Yang--Mills theory with a well-known result 
of 't~Hooft's on confinement \cite{thooft} one obtains a scheme -- the 
Dualized Standard Model -- which purports to answer some of these puzzling 
questions \cite{physcons}.

This paper reviews the theoretical basis for the scheme, while our other
 paper \cite{app97b} in the same volume reviews its physical consequences.

\section{A first look at duality}

It is well-known that electromagnetism is invariant under the
interchange $E\to H$, $H \to -E$, which can be expressed equivalently
as  a symmetry under the Hodge star operation on the field tensor
\begin{equation}
\mbox{}^*\! F_{\mu\nu} = -\half \epsilon_{\mu\nu\rho\sigma}
F^{\rho\sigma}.
\label{starf}
\end{equation}
When there are charges present, then this duality interchanges
electric and magnetic charges.

Let us take one of the Maxwell equations:
\begin{equation}
\partial_\mu \mbox{}^*\! F^{\mu\nu} =0.
\label{maxwell}
\end{equation}
Using Gauss' theorem, it is easy to see that (\ref{maxwell}) is
equivalent to the absence of magnetic monopoles.  This is the physical
content of (\ref{maxwell}).  Using the Poincar\'e lemma\footnote{This
particularly simple example of Poincar\'e lemma can easily be seen by a
direct construction of the gauge potential $A_\mu$.}, it is also easy
to see that (\ref{maxwell}) is equivalent to the existence of a gauge
potential $A_\mu$ such that $F_{\mu\nu}$ is its curl:
\begin{equation}
F_{\mu\nu} = \partial_\nu A_\mu - \partial_\mu A_\nu.
\label{curla}
\end{equation}
This is the geometric content of (\ref{maxwell}).  Notice that both
conditions are necessary and sufficient.  This situation can be
schematically represented as:
\begin{equation}
\underbrace{A_\mu {\rm \ exists}}_{\rm geometry}\ \  \stackrel{\rm
Poincar\acute{e}}{\Longleftrightarrow}\ \  \partial_\mu \mbox{}^*\! F^{\mu\nu}
=0\ \  \stackrel{\rm Gauss}{\Longleftrightarrow} \ \ \underbrace{\rm no\
magnetic\ monopoles}_{\rm physics}
\label{abscheme}
\end{equation}

The dual of eq. (\ref{maxwell}) is:
\begin{equation}
\partial_\mu F^{\mu\nu} = 0,
\label{maxwella}
\end{equation}
which is satisfied where there are no electric sources.  By the same line
of argument as for (\ref{abscheme}) we deduce that where (\ref{maxwella}) 
is satisfied, there exists also a dual potential $\tilde{A}_\mu$ such that:
\begin{equation}
\mbox{}^*\! F_{\mu\nu}=\partial_\nu \tilde{A}_\mu - \partial_\mu \tilde{A}_\nu,
\label{curlat}
\end{equation}
so that a symmetry is established under the * operation, that is
\begin{equation}
\underbrace{\tilde{A}_\mu {\rm \ exists}}_{\rm geometry}\ \ \Longleftrightarrow\ \  \partial_\mu F^{\mu\nu}
=0\ \  \Longleftrightarrow \ \ \underbrace{\rm no\
electric\ sources}_{\rm physics}
\label{dabscheme}
\end{equation}
This is the
celebrated electric-mangetic duality.

For nonabelian gauge theories, however, the picture is totally different.
Using the covariant derivative $D_\mu=\partial_\mu - ig [A_\mu,\ ]$,
we still have the analogue of (\ref{maxwell}):
\begin{equation}
D_\mu \mbox{}^*\! F^{\mu\nu} =0,
\label{bianchi}
\end{equation}
which is usually known as the Bianchi identity.  However, since there
is no nonabelian analogue to Gauss' theorem, i.e.\ in this case there
is no satisfactory way of converting a volume integral into a surface
integral, (\ref{bianchi}) has {\em nothing} to say about the existence
or otherwise of the nonabelian analogue of the magnetic monopoles.  In 
fact, even the concept of flux is lost so that one has to give an entirely 
new kind of definition to a nonabelian monopole.  Furthermore, although 
(\ref{bianchi}) holds identically for any tensor $F_{\mu\nu}$ which is 
the covariant curl of a potential $A_\mu$, the converse is false.  In
fact, on can hardly formulate the converse given that the covariant
derivative $D_\mu$ has to involve the potential $A_\mu$.  This 
means that the above diagram (\ref{abscheme}), so significant in the 
abelian case, has hardly any content in the nonabelian case:
\begin{equation}
A_\mu {\rm \ exists} \Longrightarrow D_\mu \mbox{}^*\! F^{\mu\nu} =0
\ \ \stackrel{?}{\cdots} \ \ ?
\label{noscheme}
\end{equation}
Further, there need not exist a dual potential related to $\mbox{}^*\!
F_{\mu\nu}$ in the same way as $A_\mu$ is related to $F_{\mu\nu}$.  In 
fact, Gu and Yang \cite{guyang} constructed some explicit counter-examples
of potentials $A_\mu$ which satisfy $D_\mu  F^{\mu\nu} = 0$ (and of course
$D_\mu \mbox{}^*\! F^{\mu\nu} = 0$) but no $\tilde{A}_\mu$ exists for
which $\mbox{}^*\! F_{\mu\nu}$ is its covariant curl.
So we have also the `would-be' analogue of (\ref{dabscheme}):
\begin{equation}
\tilde{A}_\mu {\rm \ exists} \ \  \stackrel{{\rm Gu-Yang}}
{\not\!\!\!\Longleftarrow}\ \  D_\mu F^{\mu\nu}=0\ \ \stackrel{{\rm
Yang-Mills}}{\Longleftrightarrow}\ \ {{\rm no\ electric\ sources}}
\end{equation}
Hence we see clearly that the nonabelian theory is not symmetric under
the Hodge star, 
as the abelian theory is. 

However, this does not mean necessarily that there is no nonabelian
generalization to duality.  Indeed, it was shown in \cite{prenabdual,ymduality}
that there is a generalized dual transform under which nonabelian theory is 
invariant.  This generalized transform (A) reduces to the usual star operation
(\ref{starf}) in the abelian case, but (B) does not do so in general in the 
nonabelian case, as it must not because of the Gu-Yang counter-examples
\cite{guyang}.

\section{Nonabelian monopoles and loop space}

Nonabelian duality is closely connected to the concept of nonabelian 
monopoles, which in turn is best expressed in the language of loop space.
We shall therefore first recall, in this section, some old results 
\cite{annals,gaugebook} on these topics, partly to introduce the notation.

Let us first recall the general definition of a monopole in a gauge theory
whether abelian or not.  Let $G$ be the gauge group.  Then a (magnetic) 
monopole is defined as the class of closed curves in $G$ \cite{monopoledef}.  
Two curves are in the same class if they can be continuously deformed into 
each other.  For example, if $G=U(1)$, then the monopole charge is given by
an integer---this is the original magnetic case.  If $G=SO(3)$, then
the monopole charge is a sign: $\pm 1$.  For the gauge group of the
Standard Model, which is our main conern, and which is strictly
speaking $SU(3) \times SU(2) \times U(1)/{\bf Z}_6$ and not 
$SU(3) \times SU(2) \times U(1)$ as usually written, the monopole
charge is again an integer $n$.  In this case, a monopole of charge
$n$ carries (a) a dual colour charge $\zeta=e^{2\pi n i/3}$, (b) a
dual weak isospin charge $\eta=(-1)^n$, and (c) a dual weak
hypercharge $\tilde{Y}=\frac{2n\pi}{3g_1}$, where $g_1$ is the weak
hypercharge coupling \cite{unifcharge}.

The monopole charge thus defined is quantized and conserved.  But how
does one express it in an equation?  We found that we can do so using
Polyakov's loop space formulation of gauge theory
\cite{polyloop,annals}.

Let $\xi^\mu (s),\ s=0 \to 2\pi$, be a closed curve in spacetime
beginning and ending in a fixed point $\xi^\mu (0)=\xi^\mu (2\pi) =
x^\mu_0$.  Then the phase factor or Wilson loop or holonomy
\cite{wyphase} is the following loop-dependent but gauge-invariant
element of the gauge group $G$:
\begin{equation}
\Phi [\xi] =P_s \exp \left( ig \int_0^{2\pi} A_\mu (\xi(s))
\dot{\xi}^\mu (s) ds \right),
\label{phasefact}
\end{equation}
where $P_s$ means path-ordering with respect to $s$.  From this we can
define the `loop space connection'
\begin{equation}
F_\mu [\xi |s] = \tfrac{i}{g} \Phi^{-1} [\xi]\,\delta_\mu (s) \Phi
[\xi]
\label{loopconn}
\end{equation}
and the corresponding `loop space curvature'
\begin{equation}
G_{\mu\nu} [\xi|s] = \delta_\nu F_\mu [\xi|s] - \delta_\mu F_\nu
[\xi|s] +ig [F_\mu [\xi|s],F_\nu [\xi|s]],
\label{loopcurv}
\end{equation}
where $\delta_\mu (s)$ denotes the loop derivative at the point $s$
on the loop.

With this apparatus, one can first of all write down for example an
$SO(3)$ monopole of charge $-1$:
\begin{equation}
G_{\mu\nu} [\xi|s] = \kappa, \quad \exp i\pi \kappa =-1.
\label{gmunu}
\end{equation}
Secondly, what is more important, one can prove the so-called extended 
Poincar\'e lemma \cite{annals}, which states that, apart from some minor 
technical conditions, the vanishing of the loop curvature is equivalent to the
existence of a local gauge potential $A_\mu$ giving rise to $G_{\mu\nu}
[\xi|s]$ in the above manner.

Thus we can now replace the contentless (\ref{noscheme}) with the true
nonabelian analogue of (\ref{abscheme}):
\begin{equation}
\underbrace{A_\mu {\rm \ exists}}_{\rm geometry} 
\ \ \Longleftrightarrow\ \  G_{\mu\nu} =0
 \ \ \Longleftrightarrow\ \  \underbrace{\rm no\
magnetic\ monopoles}_{\rm physics}
\label{nabscheme}
\end{equation}
once again linking geometry to physics via a simple condition.

\section{Nonabelian duality}

Just as we sought a nonabelian version (\ref{nabscheme}) of (\ref{abscheme}), 
we now seek to generalize the notion of duality suitable for the nonabelian 
case.  We recall that the abelian duality transformation * satisfies the 
following two conditions:
\begin{enumerate}
\item[(I)] It is its own inverse apart from a sign: $\mbox{}^*\! (\mbox{}^*\! 
    F_{\mu\nu}) = - F_{\mu\nu}$,
\item[(II)] It interchanges electricity and magnetism: $e \longleftrightarrow
    \tilde{e}$.
\end{enumerate}
We thus look for a generalized duality transformation for a nonabelian 
gauge theory which satisfies (I) and (II), requring that it (A) reduces to * 
in the abelian case but (B) does not do so in general in the nonabelian case.

First, we need to make clear what is meant by (II) in a nonabelian theory.
We recall that for the abelian theory, in the `electric' description in 
terms of $A_\mu$, an electric charge is a {\it source} represented by a 
nonvanishing current on the right-hand side of (\ref{maxwella}), while a 
magnetic charge is a {\it monopole} which in terms of $A_\mu$ is topological 
in origin but also representable by a nonvanishing dual current on the 
right-hand side of (\ref{maxwell}).  Hence, for a nonabelian theory, in 
the `electric' description in terms of $A_\mu$, an electric charge should 
also be a {\it source} represented by nonvanishing current on the left of:
\begin{equation}
D_\mu F^{\mu\nu} = j^\nu,
\label{yangmills}
\end{equation}
while a magnetic charge should be a {\it monopole} represented, by virtue
of (\ref{nabscheme}), by a nonvanishing loop curvature $G_{\mu\nu}$.

To write down the generalized duality transform, introduce the following 
set of variables \cite{ymduality}:
\begin{equation}
E_\mu [\xi|s] = \Phi_\xi (s,0) F_\mu [\xi|s] \Phi^{-1}_\xi (s,0),
\label{evariab}
\end{equation}
where
\begin{equation}
\Phi_\xi (s_1,s_2) = P_s \exp \left( ig \int_{s_1}^{s_2} A_\mu
(\xi(s)) \dot{\xi}^\mu (s) ds \right).
\label{paratrans}
\end{equation}
We see immediately that the $E$ variables are the $F$ variables
parallely transported by (\ref{paratrans}).  
%They can be represented schematically as in Fig. 2.
It is clear that $E_\mu [\xi|s]$ depends only on a segment of the loop
$\xi(s)$ around $s$, and is therefore a `segmental' variable rather
than a full `loop' variable.  In the limit that the segment shrinks to
a point, we have
\begin{equation}
E_\mu [\xi|s] \longrightarrow F_{\mu\nu}(\xi(s)) \dot{\xi}^\nu (s).
\label{elimit}
\end{equation}
However, the limit (\ref{elimit}) must not be taken before other loop
operations such as loop differentiation are performed, as these loop
operations do require at least a segment of loop on which to operate.

It is not too difficult to show that the variables $E_\mu [\xi|s]$
constitute an equivalent set of variables to $F_\mu [\xi|s]$.  Using
these, we can now define the duality transform by
\begin{eqnarray}
\lefteqn{\omega^{-1} (\eta(t)) \tilde{E_\mu} [\eta|t] \omega
(\eta(t))=}  \nonumber \\
&& - \frac{2}{\bar{N}} \epsilon_{\mu\nu\rho\sigma} \dot{\eta}^\nu (t)
\int \delta \xi ds E^\rho [\xi|s] \dot{\xi}^\sigma (s) \dot{\xi}^{-2}
(s) \delta (\xi(s)-\eta(t)).  \label{duality}
\end{eqnarray}

At first sight, this is quite unlike (\ref{starf}).  However, if we regard 
the loop dependence of $E^\rho [\xi|s]$ as a continuous index, then the 
loop integral on the right is nothing but saturating indices, just like 
the summation on the right hand side of (\ref{starf}).  By (\ref{elimit}) 
we see that it is reasonable that the tangents $\dot{\xi}^\sigma (s)$ and 
$\dot{\eta}^\nu (t)$ should occur.  The factor $\bar{N}$ is an (infinite) 
normalization constant inherent in doing the functional integral.  One 
novel ingredient is the {\em local} quantities $\omega (x)$ on the left hand
side.  For concreteness, let us take $G=SU(3)$.  Then $\omega$ is a $3
\times 3$ unitary matrix which represents the change from the frame in
internal colour space with respect to which $E_\mu$ is defined to the
frame in internal dual colour space with respect to which $\tilde{E}_\mu$ 
is defined.  Such a change in frame is necessary to balance the two 
sides of eq. (\ref{duality}) since $E_\mu$ is `electrically' charged
but `magnetically' neutral, transforming thus only under $SU(3)$, not
under its dual $\widetilde{SU}(3)$ (see the last section for a discussion
of dual gauge symmetries), while for $\tilde{E}_\mu$, the reverse holds.
In the abelian case, the factors $\omega^{-1}$ and $\omega$ commute 
through and cancel, so that there we do not see this feature.  Moreover, 
we do {\em not} always have the freedom by gauge transformations to set 
$\omega=1$ everywhere, because in the presence of charges either $E$ or 
$\tilde{E}$ (or both) has to be patched\footnote{This is similar to the 
case of the electric potential $A_\mu$ in the presence of a magnetic 
monopole, requiring either patching or equivalently the Dirac string.}, 
so that $\omega$ may have to be patched also.  It thus takes on some 
dynamical properties and, as can be seen in our companion paper \cite{app97b}, 
the rows or columns of the matrix $\omega$ can even be interpreted as 
the vacuum expectation values of Higgs fields.  As such, they play a 
crucial role in the Dualized Standard Model.

Coming back to the duality transform (\ref{duality}), we note that it 
has been constructed specifically in such a way as to satisfy the 
condition (I) above and (A) to reduce to the Hodge * in the abelian 
case yet (B) without doing so for the general nonabelian case 
\cite{ymduality}.  Furthermore, it was shown there that it satisfies 
also the condition (II) by the following chain of arguments.  As known 
already to Polyakov \cite{polyloop}, a {\it source} in the `electric' 
description in terms of $A_\mu$ can be represented in his loop notation 
of (\ref{loopconn}) as nonvanishing loop divergence $\delta^\mu(s) 
F_\mu[\xi|s] \neq 0$, which by the relation (\ref{evariab}) can also 
be expressed as nonvanishing loop divergence of $E_\mu[\xi|s]$, namely 
$\delta^\mu E_\mu[\xi|s] \neq 0$.  The duality transform (\ref{duality}), 
however, is so constructed that a nonvanishing loop divergence for 
$E_\mu$ gives a nonvanishing `loop curl' for the dual variable 
$\tilde{E}_\mu$, i.e. $\delta_\nu(t) \tilde{E}_\mu[\eta|t] - \delta_\mu(t) 
\tilde{E}_nu[\eta|t] \neq 0$.  Further, using (\ref{evariab}) again, but 
now for $\tilde{E}_\mu$, it is seen that a nonvanishing `curl' for 
$\tilde{E}_\mu$ means nonvanishing loop curvature $\tilde{G}_{\mu\nu}$, 
or in other words, by the dual of (\ref{nabscheme}), a {\it monopole} in 
the `magnetic' description.  Hence, we have that a {\it source} in the
`electric' description is a {\it monopole} in the magnetic description.
Moreover, because of (I), the converse is also true, namely that a 
`magnetic' {\it source} is the same as an `electric' {\it monopole}.  
This then is the nonabelian generalization of (II) as desired.

For a pure Yang-Mills theory with neither sources nor monopoles, then it 
follows by (\ref{nabscheme}) that both the potential $A_\mu$ and the dual 
potential $\tilde{A}_\mu$ exist, substantiating thus the claim that the 
pure theory is symmetric under the dual transform (\ref{duality}).  For 
the situation with sources and monopoles around, however, some more tools 
are needed.

\section{Dynamics and the Wu--Yang criterion}

In abelian theory, the equations of motion governing the dynamics of a 
charge, whether electric or magnetic, can be derived from its topological
definition as a monopole by the Wu--Yang criterion \cite{wycriter}.  For 
concreteness, consider first a magnetic charge regarded as a monopole
in the electric description in terms of $A_\mu$.  Instead of the usual 
minimally coupled action, one starts with the free field plus free 
particle action, which one varies under the constraint that there exists 
a magnetic monopole.  Introducing a Lagrange multiplier $\lambda_\mu$ for
the constraint, we have
\begin{equation}
{\cal A} = -\frac{1}{4} \int F_{\mu\nu} F^{\mu\nu} - \int \bar{\psi}
(i \partial_\mu \gamma^\mu -m)\psi + \int \lambda_\mu (\partial_\nu 
\mbox{}^*\! F^{\mu\nu} + 4\pi j^\mu),
\label{action}
\end{equation}
where the magnetic current $j^\mu$ is given by
\begin{equation}
j^\mu= \tilde{e} \bar{\psi} \gamma^\mu \psi,
\label{current}
\end{equation}
and $\tilde{e}$ is the magnetic coupling related to the usual electric 
coupling $e$ by the Dirac quantization condition
\begin{equation}
e \tilde{e} =2\pi.
\label{dirac}
\end{equation} 
Here we have assumed the monopole to be a Dirac particle, but we could
equally have formulated the procedure classically.  Varying with
respect to $F_{\mu\nu}$ we get
\begin{equation}
\partial_\mu F^{\mu\nu} =0,
\label{dmuf}
\end{equation}
which is equivalent by (\ref{abscheme}) and duality to the existence
of a dual potential $\tilde{A}_\mu$.  In fact we have
\begin{equation}
\tilde{A}_\mu = 4\pi \lambda_\mu,
\label{duala}
\end{equation}
with
\begin{equation}
\mbox{}^*\! F_{\mu\nu}=\partial_\nu \tilde{A}_\mu -
\partial_\mu \tilde{A}_\nu.
\end{equation}
Varying with respect to $\bar{\psi}$ we get
\begin{equation}
(i \partial_\mu \gamma^\mu -m) \psi = -\tilde{e} \tilde{A}_\mu 
\gamma^\mu \psi.   \label{psieq}
\end{equation}
Together with the constraint
\begin{equation}
\partial_\mu \mbox{}^*\! F^{\mu\nu} =-4\pi \tilde{\jmath}^\mu
\label{constraint}
\end{equation}
equations (\ref{dmuf}) and (\ref{psieq}) constitute the equations of
motion for the field--monopole system \cite{wycriter}.

The argument can be repeated for electric charges by regarding them as
monopoles in the magnetic description in terms of $\tilde{A}_\mu$.  The 
constraint is then given by
\begin{equation}
\partial_\mu F^{\mu\nu} =-4 \pi j^\mu,
\label{ecurrent}
\end{equation}
yielding instead the usual Maxwell and Dirac equations, i.e.\ exactly the
duals of (\ref{dmuf}) and (\ref{psieq}).  One concludes therefore that
electromagnetism is dual symmetric even in the presence of charges.

\section{Dynamics of nonabelian charges}

We wish next to extend the argument to nonabelian Yang-Mills theory
using the formalism developed above.  Again we shall use the Wu--Yang 
criterion to study the dynamics of nonabelian charges, regarding them 
as monopoles.  In loop variables, the free field action is
\begin{equation}
{\cal A}_F=-\frac{1}{4\pi \bar{N}} \int \delta \xi ds {\rm Tr} (E_\mu
E^\mu) \dot{\xi}^{-2}.     \label{actionf}
\end{equation}
The free (Dirac) particle action is as before
\begin{equation}
{\cal A}_M = \int \bar{\psi} (i \partial_\mu \gamma^\mu -m) \psi.
\label{actionm}
\end{equation}
The constraint that there is a monopole is
\begin{equation}
\partial_\nu E_\mu - \partial_\mu E_\nu = -4\pi J_{\mu\nu},
\label{actionj}
\end{equation}
where the magnetic current $J_{\mu\nu}$ has the form
\begin{equation}
J_{\mu\nu} [\xi|s]=\tilde{g} \epsilon_{\mu\nu\rho\sigma} (\bar{\psi}
\omega \gamma^\rho t^i \dot{\xi}^\sigma \omega^{-1} \psi) t_i,
\label{jmunu}
\end{equation}
and $t_i$ is a generator in the relevant representation of $G$.
The monopole charge is originally given as a nonvanishing loop
curvature $G_{\mu\nu}$, which is the loop covariant curl of $F_\mu$.
However, as already mentioned above, it can be shown that by going over 
to the variables $E_\mu$ by (\ref{evariab}), the loop covariant curl 
becomes simply the loop curl; hence the constraint (\ref{actionj}). 

The full action
\begin{equation}
{\cal A}={\cal A}_F + {\cal A}_M + \int \delta \xi ds {\rm Tr}
(W^{\mu\nu} (\partial_\nu E_\mu - \partial_\mu E_\nu +4\pi J_{\mu\nu}))
\end{equation}
is then varied with respect to the variables $E_\mu [\xi|s]$ and
$\bar{\psi} (x)$, giving respectively
\begin{eqnarray}
\delta^\mu (s) E_\mu [\xi|s] & = & 0 \label{divge} \\
(i \partial_\mu \gamma^\mu -m) \psi (x) & = & -\tilde{g} \tilde{A}_\mu
\gamma^\mu \psi(x),  \label{nonpsieq}
\end{eqnarray}
where the dual potential $\tilde{A}_\mu$ is given by the Lagrange
multiplier $W^{\mu\nu}$, in analogy to (\ref{duala}):
\begin{equation}
\tilde{A}_\mu (x)= 4\pi \epsilon_{\mu\nu\rho\sigma} \int \delta \xi ds
\omega(\xi(s)) W^{\rho\sigma} [\xi|s] \omega^{-1} (\xi(s))
\dot{\xi}^\nu \dot{\xi}^{-2} \delta (\xi(s)-x).
\end{equation}
As already noted above, (\ref{divge}) is equivalent to the usual Yang--Mills
source-free equation
\begin{equation}
D^\mu F_{\mu\nu} =0.   \label{ym}
\end{equation}

To study the dynamics of nonabelian electric charges, we start from
Yang--Mills equation (\ref{ym}) with a nonvanishing right hand side.
This implies that
\begin{equation}
\delta^\mu (s) E_\mu [\xi|s] \not= 0,
\end{equation} 
which in turn implies
\begin{equation}
\delta_\nu \tilde{E}_\mu - \delta_\mu \tilde{E}_\nu \not= 0.
\end{equation}
But this is the condition that signals the occurrence of a monopole
of the $\tilde{E}_\mu$ field (cf.\ (\ref{actionj})).  Since the free
field action (\ref{actionf}) can equally be expressed in terms of the
dual variables:
\begin{equation}
{\cal A}_F = \frac{1}{4\pi \bar{N}} \int \delta \xi ds {\rm Tr}
(\tilde{E}_\mu \tilde{E}^\mu) \dot{\xi}^{-2},
\end{equation}
we can easily derive, by imposing the appropriate constraint
\begin{equation}
\delta_\nu \tilde{E}_\mu - \delta_\mu \tilde{E}_\nu= -4\pi J_{\mu\nu}
\end{equation}
with an expression for the current similar to (\ref{jmunu}), the equations 
of motion of nonabelian electric charges as: 
\begin{eqnarray}
\delta^\mu (s) \tilde{E}_\mu [\xi|s] & = & 0 \\
(i \partial_\mu \gamma^\mu -m) \psi (x) & = & -g A_\mu \gamma^\mu \psi(x).
\end{eqnarray}

We see that the equations of motion for the nonabelian electric charge
are exactly the duals of those given above for the nonabelian magnetic
charge, namely (\ref{divge}) and (\ref{nonpsieq}).  Hence we conclude 
that, as claimed, the dynamics is indeed symmetric under the generalized 
duality transform (\ref{duality}) even in the presence of charges, just 
as in the abelian case.

\section{Remarks and conclusion}

We have presented the salient features of nonabelian duality, without
supplying many details.  A few remarks, therefore, are in order.

Firstly, since both the potentials $A_\mu (x)$ and $\tilde{A}_\mu (x)$
are local spacetime variables, it may be tempting to speculate that
the duality transform (\ref{duality}) itself could perhaps be
formulated entirely in terms of local spacetime variables rather than
loop variables.  At present, we certainly do not know a way of doing
that.  Suppose we start with the variables $A_\mu (x)$, then by
(\ref{nabscheme}) we deduce that in the presence of monopoles $A_\mu
(x)$ cannot be everywhere defined.  If at the same time there are no
sources, then $\tilde{A}_\mu (x)$ is everywhere defined.  By duality
the existence of sources, while allowing $A_\mu (x)$ to be everywhere
defined, forces $\tilde{A}_\mu (x)$ to be undefined in certain regions
of spacetime.  This means that if there is only one type of charges
present (whether monopole or source), we may, by choosing our
variables, stick to spacetime variables only.  However, if {\rm both}
charges, or dyons, are present, then it seems that loop space
variables are inevitable.  Unfortunately, the rigorous mathematics of 
loop space analysis remains largely unexplored \cite{newman}.  For the 
work reported above, we have devised certain operational rules which 
seem to us consistent at least for the use we put them to, but the lack
for a general loop calculus is often acutely felt.  For more details, 
we refer the interested reader to \cite{ymduality} and earlier work 
cited therein.  Nevertheless, the existing operational rules already
allows one to explore Feynman diagram techniques using loop space 
variables \cite{feynloop}, which can be a first step towards building 
a full quantum field theory in these variables. 

Secondly, because of dual symmetry, a nonabelian gauge theory is not
invariant just under the usual gauge group $G$ but rather two copies
of it: $G \times \tilde{G}$.  Here we denote the group under which
$\tilde{A}_\mu$ transforms as $\tilde{G}$, although as a group it is
identical to $G$.  This makes it easier notationally and also
underlines the fact that $\tilde{G}$ has parity opposite to that of
$G$, because of the $\epsilon$-symbol in the transform (\ref{duality}).  
This extra symmetry is a direct consequence of duality, which in turn 
is inherent in any gauge theory.  That this symmetry has interesting 
physical consequences will be shown in detail in our companion article 
\cite{app97b}.  We note further that although the gauge symmetry 
is found to be doubled, the number of degrees of freedom
remains the same.  In a way not yet fully explored, the potentials
$A_\mu (x)$ and $\tilde{A}_\mu (x)$ represent the same degrees of
freedom, since the duality transform (\ref{duality}) is an equation
relating $E_\mu [\xi|s]$ and $\tilde{E}_\mu [\xi|s]$.  The situation
is even more immediately evident in the abelian case.  Under a $U(1)$
transformation $\lambda (x)$,
\begin{eqnarray}
A_\mu (x) & \mapsto & A_\mu (x) + \partial_\mu \lambda (x) \nonumber
\\
\tilde{A}_\mu (x) & \mapsto & \tilde{A}_\mu (x);
\end{eqnarray}
while under a $\widetilde{U(1)}$ transformation $\tilde{\lambda} (x)$,
\begin{eqnarray}
A_\mu (x) & \mapsto & A_\mu (x) \nonumber \\
\tilde{A}_\mu (x) & \mapsto & \tilde{A}_\mu (x) + \partial_\mu
\tilde{\lambda} (x).
\end{eqnarray}
The two phases $\lambda (x)$ and $\tilde{\lambda} (x)$ are entirely
independent.  Similarly the wave function $\psi (x)$ of an electric
charge and the wave function $\tilde{\psi} (x)$ of a  magnetic
monopole will transform under $\lambda (x)$
\begin{eqnarray}
\psi(x) & \mapsto & e^{i \lambda (x)} \psi (x) \nonumber \\
\tilde{\psi} (x) & \mapsto & \tilde{\psi} (x);
\end{eqnarray}
and under $\tilde{\lambda} (x)$
\begin{eqnarray}
\psi (x) & \mapsto & \psi (x) \nonumber \\
\tilde{\psi} (x) & \mapsto & e^{i \tilde{\lambda} (x)} \tilde{\psi}
(x).
\end{eqnarray}
However, the variables $A_\mu (x)$ and $\tilde{A}_\mu (x)$ clearly do
not represent different degrees of freedom, because their field
tensors $F_{\mu\nu} (x)$ and $\mbox{}^*\!F_{\mu\nu} (x)$ are related
by the following {\em algebraic} equation
\begin{equation}
\mbox{}^*\!F_{\mu\nu} (x) = -\half \epsilon_{\mu\nu\rho\sigma}
F^{\rho\sigma} (x).
\end{equation}
That $A_\mu$ and $\tilde{A}_\mu$ should correspond to two gauge symmetries
but yet represent the same physical degree of freedom can have very 
interesting physical consequences \cite{app97b,physcons,airshow2,airshow3}.

Thirdly, since $\tilde{A}_\mu (x)$ is a local field, we can construct
{\em its} phase factor:
\begin{equation}
\tilde{\Phi} [\xi] = P_s \exp i \tilde{g} \int_0^{2\pi} \tilde{A}_\mu
(\xi(s)) \dot{\xi}^\mu (s) ds
\label{dualphase}
\end{equation}
in complete analogy to the familiar $\Phi [\xi]$ in (\ref{phasefact}).
Now, in the famous work of 't~Hooft \cite{thooft} on confinement the 
trace of $\Phi [\xi]$ has a very important role to play as an order
parameter which he called $A(C)$, depending on the loop $C$.  Hence, by the
duality discussed above, one expects that the trace of $\tilde{\Phi} [\xi]$
in (\ref{dualphase}) will play the role of 't Hooft's disorder parameters 
$B(C)$ \cite{thooft}.  This turns out to be indeed the case.  Using 
Dirac's quantization condition 
\begin{equation}
g \tilde{g} = 4 \pi
\end{equation} 
it was shown \cite{comrel} that the traces of $\Phi$ and $\tilde{\Phi}$
does indeed satisfy the commutation relation
\begin{equation}
A(C) B(C') = B(C') A(C) \exp (2\pi i n/N)
\end{equation}
for $G=SU(N)$, as required by 't~Hooft for his order-disorder parameters
\cite{thooft}.   It follows then that we can apply 't Hooft's confinement 
result \cite{thooft} to our situation, namely that if the $G$ symmetry
is confined then the $\tilde{G}$ symmetry as defined above is broken and 
Higgsed, and vice versa.  As can be seen in our companion paper \cite{app97b},
this plays a crucial role in the Dualized Standard Model \cite{physcons}.
When applied, for example, to the confined colour group $SU(3)$, it
implies a completely broken dual colour symmetry $\widetilde{SU}(3)$ 
which may be identified with generations.  The explicit form (\ref{dualphase})
for the 't~Hooft disorder parameter $B(C)$, which up to quite recently was
known only by a somewhat abstract definition, is likely to be useful also
in the problem of confinement \cite{zeni}.

Apart from giving rise to the physical consequences reviewed in \cite{app97b}, 
ranging from masses of fermions and their mixing \cite{massckm} to 
flavour-changing neutral current decays \cite{airshow1,airshow3,fcnc} 
and very high energy cosmoc rays \cite{airshow1,airshow2,airshow3} the 
considerations above raise also some intriguing theoretical questions 
that are beginning to be asked.  For example, throughout this lecture so 
far we have been concerned only with the nonabelian genralization of 
electric--magnetic duality in a strictly non-supersymmetric context
and in exactly 4 spacetime dimensions.  We have not touched upon the 
possible extension to supersymmetry and/or higher spacetime dimensions.
This could be interesting, given the vast amount of exciting work
\cite{ssduality} which has been done in recent years following the 
seminal papers of Seiberg and Witten \cite{seibwit} on supersymmetric 
duality.  In a completely different direction, the doubling of the symmetry 
is reminiscent of complexification in geometry and particularly general
relativity.  One would like to know how this generalized duality
relates to the vast literature of self-dual fields, both in geometric
Yang--Mills theory and in general relativity, especially in the
twistor description \cite{ward,nlgrav}.  The vistas that are being
opened up are truly fascinating.

Previous collaborations with Peter Scharbach and Jacqueline Faridani
are gratefully acknowledged.

\end{document}